# Approximate Early Output Asynchronous Adders based on Dual-Rail Data Encoding and 4-Phase Return-to-Zero and Return-to-One Handshaking

P. Balasubramanian

*Abstract*—Approximate computing is emerging as an alternative to accurate computing due to its potential for realizing digital circuits and systems with low power dissipation, less critical path delay, and less area occupancy for an acceptable trade-off in the accuracy of results. In the domain of computer arithmetic, several approximate adders and multipliers have been designed and their potential have been showcased versus accurate adders and multipliers for practical digital signal processing applications. Nevertheless, in the existing literature, almost all the approximate adders and multipliers reported correspond to the synchronous design method. In this work, we consider robust asynchronous i.e. quasi-delay-insensitive realizations of approximate adders by employing delay-insensitive codes for data representation and processing, and the 4-phase handshake protocols for data communication. The 4-phase handshake protocols used are the return-to-zero and the return-to-one protocols. Specifically, we consider the implementations of 32-bit approximate adders based on the return-to-zero and return-to-one handshake protocols by adopting the delay-insensitive dual-rail code for data encoding. We consider a range of approximations varying from 4-bits to 20-bits for the least significant positions of the accurate 32-bit asynchronous adder. The asynchronous adders correspond to early output (i.e. early reset) type, which are based on the well-known ripple carry adder architecture. The experimental results show that approximate asynchronous adders achieve reductions in the design metrics such as latency, cycle time, average power dissipation, and silicon area compared to the accurate asynchronous adders. Further, the reductions in the design metrics are greater for the return-to-one protocol compared to the return-to-zero protocol. The design metrics were estimated using a 32/28nm CMOS technology.

## I. Introduction

APPROXIMATE computing [1 – 3] is emerging as an alternative to accurate computing given that various digital signal processing applications such as image, video, and audio processing etc. can tolerate minor degradation in the quality of results, which may not be noticeable due to the limitations of human perception [4 – 6], to achieve reduced design metrics. This implies that approximate results which correspond to a specified error bound are acceptable.

In computing units, arithmetic operations such as additions and multiplications are found to be responsible for a majority of the power consumption. For example, more than 70% of the power consumed by a graphics processing unit is attributed to arithmetic operations [7], and about 80% of the power consumed by a fast Fourier transform (FFT) processor is attributed to adders and multipliers [8]. The FFT and inverse FFT operations are common in the OFDM transceiver, used in a wireless communication system. Further, in a JPEG encoder or decoder, which is used for digital image processing, or in a MPEG encoder or decoder, which is used for digital video processing in multimedia applications, the discrete cosine transform (DCT) and the inverse DCT operations are common which involve additions and multiplications.

Computer arithmetic is indeed pervasive in digital signal processing, and adders and multipliers are predominant in the datapath of a digital signal processing unit. Hence, the bulk of the reported research on approximate computing has focused on the design of approximate adders and multipliers [9] [10]. However, almost all the approximate adders and multipliers reported in the literature correspond to the synchronous design method. Reference [11] is perhaps the first work that discussed the implementation of approximate quasi-delay-insensitive (QDI) asynchronous adders and evaluated their performance vis-à-vis accurate QDI asynchronous adders. The delay-insensitive dual-rail code was used for data encoding, and the 4-phase return-to-zero (RTZ) handshake protocol was used for data communication. Weak-indication and early output 32-bit approximate adders, which incorporate approximation sizes ranging from 4- to 20-bits in the least significant positions, were implemented alongside the accurate 32-bit asynchronous adders. It was observed the approximate asynchronous adders paved the way for optimization of the design metrics such as latency, cycle time, area, and average power dissipation compared to the accurate asynchronous adders. Also, it was observed the early output approximate asynchronous adders exhibit improved design metrics than the weak-indication approximate asynchronous adders.

This work builds upon [11] by implementing approximate asynchronous adders based on the 4-phase return-to-one (RTO) handshake protocol, besides the 4-phase RTZ handshake protocol. This is important since it was shown recently in [12] that the 4-phase RTO protocol could facilitate enhanced optimization of the design metrics compared to the 4-phase RTZ protocol for QDI asynchronous arithmetic circuits. In this work, we specifically consider approximate implementations of robust early output asynchronous adders for approximation sizes varying from 4- to 20-bits in the least significant positions and compare them with the accurate implementations based on the RTZ and RTO handshake protocols. We adopt the delay-insensitive dual-rail code for data encoding. The accurate and approximate early output asynchronous adder implementations are QDI. QDI [13] asynchronous circuits are the practically realizable delay-insensitive asynchronous circuits with the only exception and assumption of isochronic forks, which form the weakest compromise to delay-insensitivity. Isochronicity implies that the signal transitions are assumed to happen concurrently at all the ends of an isochronic fork. It was shown in [14] that isochronicity is realizable even in nanoelectronic circuits.

The rest of the article is organized as follows. Section II discusses the fundamentals of QDI asynchronous circuit design. Section III describes the approximate asynchronous adder architecture, and portrays the approximate asynchronous adder components based on the 4-phase RTZ and RTO handshake protocols. Section IV presents the simulation results corresponding to the accurate and approximate 32-bit asynchronous adders based on physical implementation using a 32/28nm CMOS process. Finally, Section V concludes and also suggests a direction for further work.

## II. QDI ASYNCHRONOUS CIRCUIT DESIGN

A background about QDI asynchronous circuit design is provided by describing the delay-insensitive dual-rail data encoding and the 4-phase RTZ and RTO handshaking. Also, the various types of QDI asynchronous circuits are discussed.

### A. Dual-Rail Data Encoding and 4-Phase Handshaking

The dual-rail code, also known as the 1-of-2 code, is the simplest member of the family of delay-insensitive *m*-of-*n* codes [15]. Among the *m*-of-*n* codes, the 1-of-*n* codes represent a subset and are called one-hot codes. In a 1-of-*n* code, only 1 out of *n* wires is asserted as 1 to represent the binary data. The 1-of-*n* coding scheme is said to be unordered [16] since none of the code words forms a subset of another code word. Also, the 1-of-*n* coding scheme is said to be complete [17] if all the *n* unique code words are utilized to encode the specified binary inputs.

When adopting the 4-phase RTZ protocol [18], and as per the dual-rail code, a single-rail binary input W is encoded using two wires as say, W1 and W0. W = 1 is represented by W1 = 1 and W0 = 0, and W = 0 is represented by W1 = 0 and W0 = 1. Note that W1 and W0 cannot assume 1 concurrently as it is illegal and invalid since the coding scheme will become unordered. However, W1 and W0 can assume 0 concurrently and it is called the spacer. Hence, when utilizing the 4-phase RTZ protocol for data communication, and as per the dual-rail code, the data is specified by either W1 or W0 assuming 1 and the other assuming 0, and the condition of both W1 and W0 assuming 0 is called the spacer. Thus the spacer is an all-zero in the case of the 4-phase RTZ protocol.

On the other hand, when adopting the 4-phase RTO protocol [19], and as per the dual-rail code, a single-rail binary input W is encoded using two wires as say, W1 and W0, where W = 1 is represented by W1 = 0 and W0 = 1, and W = 0 is represented by W1 = 1 and W0 = 0. Note that W1 and W0 cannot assume 0 concurrently. However, W1 and W0 can assume 1 concurrently and is referred to as the spacer. Hence, when employing the 4-phase RTO handshake protocol for data communication, and as per the dual-rail code, the data is specified by either W1 or W0 assuming binary 0 and the other assuming binary 1, and the condition of both W1 and W0 assuming binary 1 is called the spacer. Hence, there is an all-one spacer in the case of the 4-phase RTO protocol.

A QDI asynchronous circuit stage that employs the delay-insensitive dual-rail code for data representation and processing and a 4-phase RTZ or RTO handshake protocol for data communication is shown in Fig 1. As the name implies, a 4-phase handshake protocol consists of four phases which will be explained with reference to Fig 1 by assuming the dual-rail encoded data. Nevertheless, this explanation would be applicable for data represented using any 1-of-*n* code. We first describe the 4-phase RTZ handshaking, followed by the 4-phase RTO handshaking.

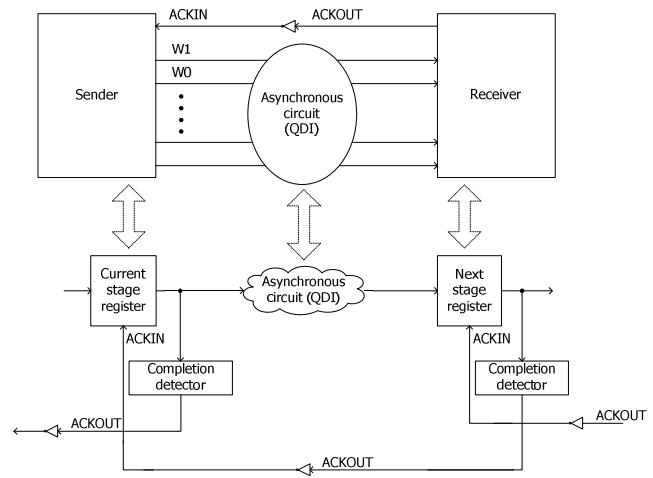

Fig 1 A QDI asynchronous circuit stage correlated with the sender-receiver analogy

According to the 4-phase RTZ protocol, in the first phase, the dual-rail data bus shown in Fig 1 which is specified by (W1, W0) etc. is in the spacer state and so ACKIN is 1. The sender transmits a code word i.e. data and this results in rising signal transitions from 0 to 1 on anyone of the corresponding dual rails of the entire dual-rail data bus. In the second phase, the receiver receives the code word sent, and then it drives ACKOUT to 1. In the third phase, the sender waits for ACKIN to become 0 and then resets the entire dual-rail data bus (i.e. spacer). Subsequently, in the fourth phase, after an unbounded time duration, which is deemed finite and positive, the receiver drives ACKOUT to 0 i.e. ACKIN becomes 1. With this, one data transaction is said to be completed, and the asynchronous circuit stage is allowed to commence the next data transaction. Therefore, the application of input data follows the sequence: data-spacer-data-spacer, and so forth.

According to the 4-phase RTO protocol, in the first phase, ACKIN is 1. The sender transmits the spacer and this results in rising signal transitions on the entire dual-rail data bus. In the second phase, the receiver receives the spacer sent, and it drives ACKOUT to 1. In the third phase, the sender waits for ACKIN to become 0 and then sends the data by resetting anyone of the corresponding rails of the entire dual-rail data bus. Subsequently, in the fourth phase, after an unbounded time duration, which is deemed finite and positive, the receiver drives ACKOUT to 0 i.e. ACKIN becomes 1. One data transaction is now said to be completed, and the asynchronous circuit stage is permitted to commence the next data transaction. Thus the application of input data follows the sequence: spacer-data-spacer-data, and so forth.

### B. Types of QDI Asynchronous Circuits

QDI asynchronous circuits are generally categorized as strong-indication [20] [21], weak-indication [20] [22], and early output [23] [24] types. Indication means providing acknowledgment for the receipt of the primary inputs through the primary outputs. This is accomplished by ensuring that indication is also provided by the intermediate outputs [18]. With respect to the asynchronous circuit stage shown in Fig 1, the indication mechanism may be local or global [25] [26]. The indication mechanism is called local if the asynchronous circuit by itself is capable of acknowledging the receipt of all the primary inputs. The indication mechanism is called global if the asynchronous circuit stage on the whole indicates the receipt of all the primary inputs in conjunction with the asynchronous circuit present within it. The input-output timing behavior of strong-indication, weak-indication, and early output asynchronous circuits is illustrated by a representative timing diagram shown in Fig 2.

A strong-indication asynchronous circuit starts data processing to produce the required primary outputs only after receiving all the primary inputs whether they are data or spacer. A weak-indication asynchronous circuit could start data processing and produce some of the primary outputs after receiving just a subset of the primary inputs. Nonetheless, the production of at least one primary output is delayed till the last primary input is received. An early output asynchronous circuit could start data processing and produce all the primary outputs after receiving just a subset of the primary inputs. If all the primary outputs are produced after receiving the data on a subset of the primary inputs, the early output asynchronous circuit is said to be of early set type. On the other hand, if the spacer is produced on all the primary outputs after receiving the spacer on a subset of the primary inputs, the early output asynchronous circuit is said to be of early reset type. The early set and reset properties of early output asynchronous circuits are depicted through the violet and orange ovals in dotted lines in Fig 2. Among the different timing models, the strong-indication is the most restrictive and the early output is more relaxed. The early output asynchronous circuits could pave the way for enhanced optimizations of the design metrics compared to strong-indication or weak-indication circuits, and this has been demonstrated through many works in the literature [24] [27 – 32].

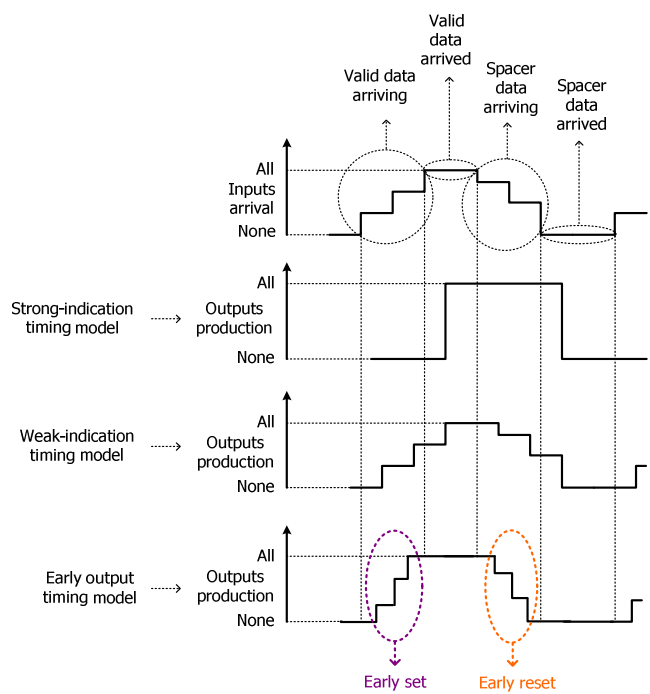

Fig 2 Strong-indication, weak-indication, and early output timing models for QDI asynchronous circuits

In a QDI asynchronous circuit, any transition on the primary inputs are required to propagate monotonically i.e. unidirectionally throughout the entire circuit depth from the primary inputs to the primary outputs with no unacknowledged signal transition on any intermediate gate output [33]. For indication, the signal transitions should either monotonically increase from binary 0 to 1, or monotonically decrease from binary 1 to 0 throughout the entire circuit. For data represented using the dual-rail code and communicated based on the 4-phase RTZ handshaking, when data are supplied the transitions would monotonically increase and for the application of spacer the transitions would monotonically decrease throughout the circuit depth. On the other hand, for data represented using the dual-rail code and communicated based on the 4-phase RTO handshaking, when the spacer is supplied the transitions would monotonically increase and for the application of data, the transitions would monotonically decrease throughout the circuit depth. An unacknowledged signal transition on an intermediate gate output is termed as gate orphan, which is to be avoided in a QDI asynchronous circuit. The issue of gate orphan has been clearly explained through diverse scenarios in [34 – 38].

Care should be taken to ensure that any logic transformation or optimization performed in a QDI asynchronous circuit conforms to the safe QDI logic decomposition principles [39]. This is because indication and robustness go hand-in-hand in QDI asynchronous circuits, and any arbitrary decomposition of logic gate(s) might give rise to gate orphan(s) which could potentially affect the robustness of a QDI circuit. Moreover, resolving the gate orphan(s) is non-trivial and may require extensive timing analysis [40] and perhaps additional timing assumptions which could complicate the physical realization of a QDI circuit. Further, if gate orphans are left unresolved, they may become problematic to a QDI circuit or system operation [41] [42], and might even cause a stall.

### III. APPROXIMATE ASYNCHRONOUS ADDERS

An $n$-bit ripple carry adder (RCA) is realized by cascading ($n - 1$) full adders with a least significant half adder. The half adder adds an augend and an addend input and produces the sum and carry overflow outputs. On the other hand, the full adder [43 – 45] adds an augend and an addend input along with any carry input and produces the sum and carry outputs. The accurate 32-bit QDI asynchronous RCA is shown in Fig 3a, which consists of 31 full adders (FA31 to FA1) and a half adder (HA1). Note that the inputs and outputs of the accurate and approximate 32-bit asynchronous adders shown in Fig 3 are dual-rail encoded, with a 4-phase RTZ or RTO protocol used for handshaking.

The approximate 32-bit QDI asynchronous adders are shown in Figs 3b to 3f, with 4-, 8-, 12-, 16-, and 20-bits approximation incorporated in the least significant adder positions. The approximate adders shown in Fig 3 basically consist of an accurate sub-adder and an approximate sub-adder. Addition is performed accurately in the former and inaccurately in the latter. The number of bits allotted to the accurate and approximate sub-adders are clearly marked in Figs 3b to 3f. Full adders are used to produce the accurate sum bits of the accurate sub-adders, and 2-input OR gates (shown as OR1 to OR20 in Figs 3b to 3f) are used to produce the approximate sum bits of the approximate sub-adders. The most significant augend and addend bit pair of the approximate sub-adder is AND-ed and its output is supplied as the carry input to the accurate asynchronous sub-adder. If the logical product of the most significant augend and addend bit pair of an approximate sub-adder yields 1, then a carry input of 1 is supplied to the accurate sub-adder; otherwise a carry input of 0 is supplied in the dual-rail encoded form.

The approximate adders, portrayed by Figs 3b to 3f, are derived from the approximate adder architecture presented in [46] but with the exception that these approximate adders correspond to QDI asynchronous implementations. The utility of the approximate adder of [46] had been demonstrated through soft-computing applications such as a 3-layer face recognition neural network, and the hardware de-fuzzification block of a fuzzy processor.

Accurate and approximate early output 32-bit QDI asynchronous adders were realized using the standard library cells of a 32/28nm CMOS process [47]. The 2-input C-element was alone manually realized using the AO222 complex gate by incorporating feedback. The C-element is indispensable in QDI asynchronous circuit designs, and would output 0 or 1 if all its inputs are 0 or 1 respectively. However, if the inputs to a C-element are non-identical, the C-element would maintain its existing steady-state. The C-element is represented by the circle with the marking C in Fig 4.

The dual-rail full adder and half adder form the building blocks of the accurate 32-bit asynchronous adder depicted in Fig 3a, and the dual-rail full adder, half adder, 2-input AND, and 2-input OR form the building blocks of the approximate 32-bit asynchronous adders depicted in Figs 3b to 3f. All the building blocks used correspond to the early output type. The logic compositions of the dual-rail full adder, half adder, 2-input AND, and the 2-input OR are shown in Fig 4. Figs 4a, 4c, 4e and 4g show the implementations of the building blocks in adherence to the RTZ protocol, and Figs 4b, 4d, 4f and 4h show the implementations according to the RTO protocol.

The rules for transforming a logic corresponding to the RTZ protocol into that suitable for the RTO protocol, and vice-versa, have been stated and proved in [12], and the interested reader is referred to the same for details. In general, the logic transformation rules governing the conversion from RTZ to RTO, and vice-versa, are found to obey the duality principle of Boolean algebra. The duality principle states that a logic expression derived by interchanging the logical operators and the identity elements of an original logic expression also remains valid [48]. However, it is important to note that the logic transformation rules based on the duality principle, which govern the conversion between the RTZ and RTO protocols, are applicable only to the discrete and complex logic gates, and not to the C-elements. As seen in Figs 4a and 4b, the inputs to the C-elements remain unchanged when transforming a logic corresponding to the RTZ protocol into that adhering to the RTO protocol, and vice-versa.

It may be worth mentioning how the basic building blocks shown in Fig 4 are constructed. The dual-rail full adder shown in Fig 4a [24] is synthesized using the disjoint sum-of-products (DSOP) expression governing the full adder. In a DSOP equation [49] [50], the logical conjunction of any two products yields 0 i.e. the product terms are mutually orthogonal [51]. The logic rules, stated above, are applied to transform the dual-rail full adder of Fig 4a into the dual-rail full adder shown in Fig 4b. With respect to the dual-rail half adder shown in Fig 4c, the sum equations are inherently in the DSOP form. The dual-rail sum output are synthesized using single complex gates to facilitate early output, and these gates are replaced by their duals to synthesize the sum output of Fig 4d. However,

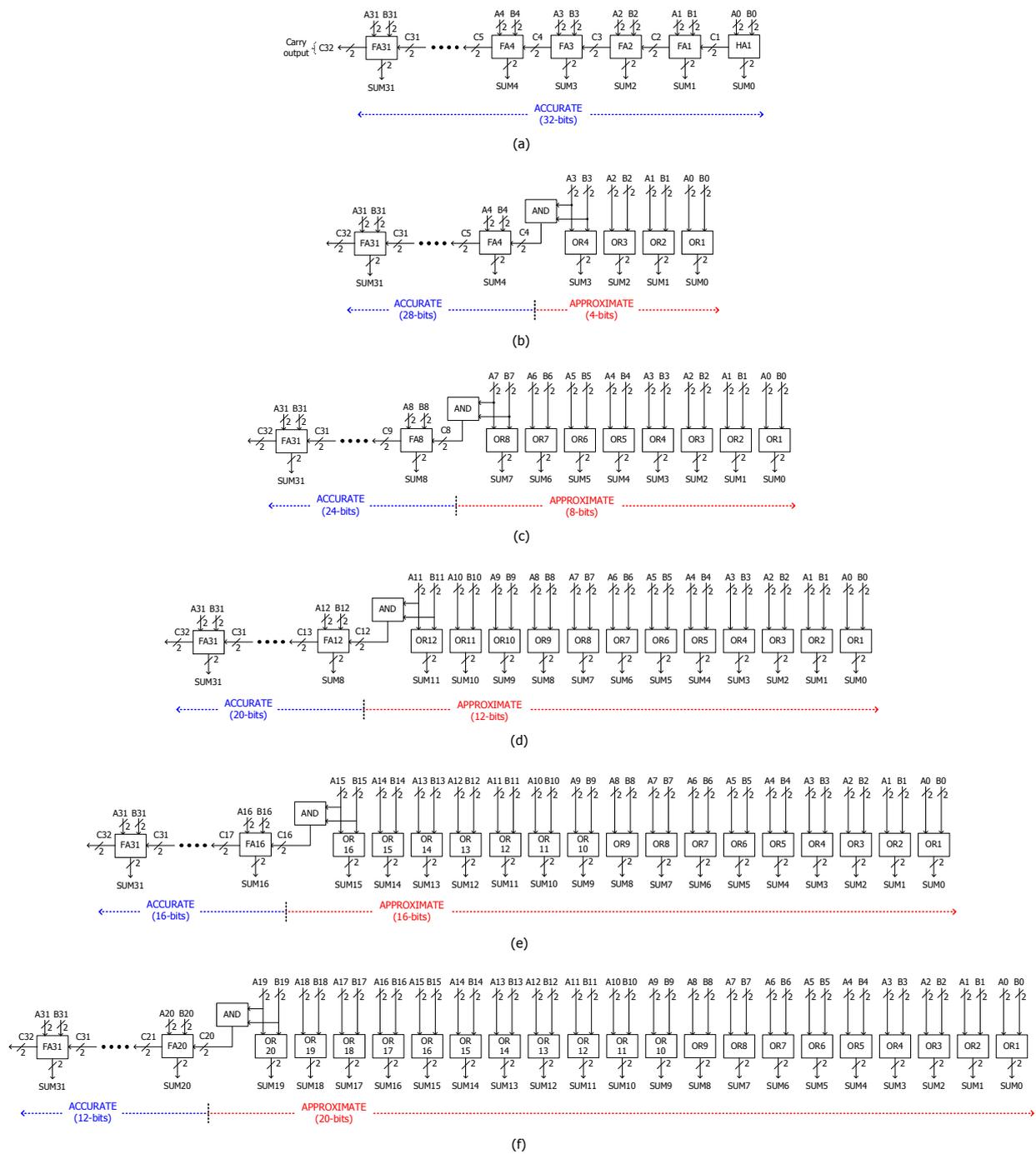

Fig 3 (a) Accurate 32-bit QDI asynchronous adder – A31 and B31 is the most significant input bit-pair, A0 and B0 is the least significant input bit-pair, SUM31 and SUM0 are the most significant and the least significant sum bits, and C32 is the carry output, which are all dual-rail encoded; Approximate QDI asynchronous adders with (b) 4-bits, (c) 8-bits, (d) 12-bits, (e) 16-bits, and (f) 20-bits approximation in the least significant the dual-rail carry output of Fig 4c is not in the DSOP form. Hence the dual-rail carry of Fig 4c is synthesized using a simple and a complex logic gate to avoid any problem of gate orphans. The duals of these gates are then used to synthesize the corresponding RTO logic equivalent, as shown in Fig 4d. The early output 2-input AND and OR logic functions are implemented as shown in [23]. The true-rail of the AND gate (Z1 in Fig 4e) and the OR gate (V1 in Fig 4g) are synthesized according to their basic logic functions. The corresponding false-rails of the AND gate and the OR gate are synthesized by complementing the true-rail outputs. The duals of the dual-rail AND gate and OR gate outputs are derived to synthesize the corresponding logic conforming to the RTO protocol, as shown in Figs 4f and 4h. The dual-rail full adder and the half adder incorporate redundant logic [52], which is implicit. This is the case with respect to both the RTZ and RTO handshake protocols. The dual-rail implementations of the AND and OR functions, however, do not feature redundant logic.

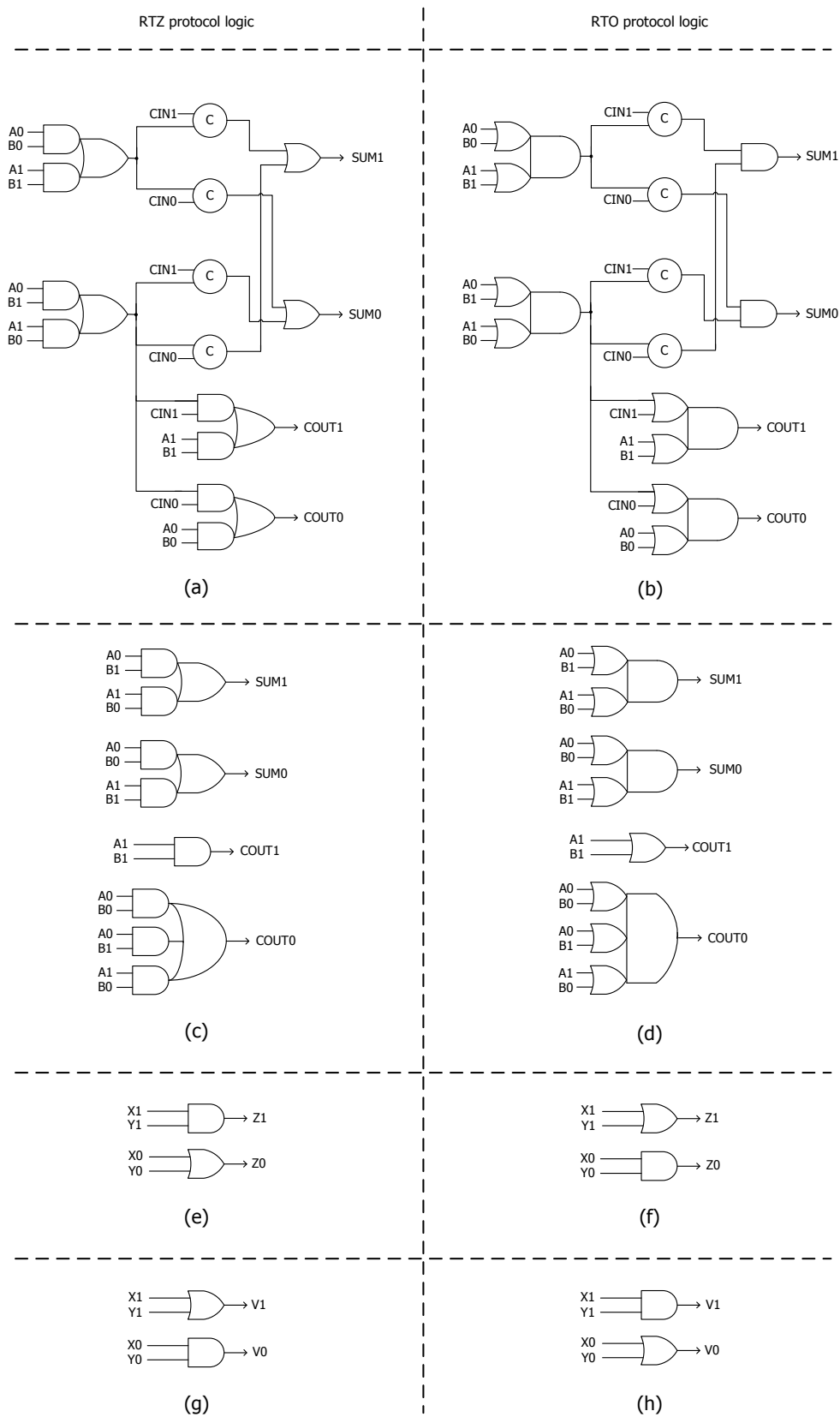

Fig 4 Early output basic building blocks. Full adder realized according to (a) RTZ protocol and (b) RTO protocol; Half adder realized according to (c) RTZ protocol and (d) RTO protocol; AND function implementation according to (e) RTZ protocol and (f) RTO protocol; OR function implementation according to (g) RTZ protocol and (h) RTO protocol

## IV. RESULTS AND DISCUSSION

Accurate and approximate 32-bit QDI asynchronous adders, which correspond to the early output type, were realized in semi-custom ASIC design style using the digital library cells of a 32/28nm CMOS process [47]. As mentioned earlier, the 2-input C-element was alone manually realized. Minimum sized cells were used uniformly for realizing all the adders. The inputs and outputs of the asynchronous adders are dual-rail encoded, and the adders conform to the RTZ and RTO handshake protocols. The approximate adders incorporate approximations ranging from 4-bits to 20-bits in the least significant positions, as shown in Fig 3.

A QDI asynchronous circuit stage, which is shown in Fig 1, consists of the asynchronous circuit, the input registers, and the completion detector. The input registers and the completion detector of the asynchronous adders corresponding to the RTZ and RTO protocols are respectively identical. Therefore, any differences between the various asynchronous adders are entirely attributed to the diversities in their logical composition (i.e. the adder logic), and this explains the reason behind the differences in the corresponding experimental results obtained. Further, this observation paves the way for a straightforward comparison of the design metrics of different asynchronous adders which correspond to the RTZ and RTO protocols post physical synthesis.

Latency, cycle time, area, and average power dissipation are the design metrics estimated for the different asynchronous adders using Synopsys tools. Here, latency generally implies forward latency which is the maximum propagation delay encountered in an asynchronous adder for the application of data. Cycle time refers to the sum of forward and reverse latencies, where the reverse latency is the maximum propagation delay encountered in an asynchronous adder for the application of spacer. In synchronous circuits, the latency specifies the rate at which new data can be input to a circuit but in QDI asynchronous circuits, the cycle time determines the rate at which new data can be input. This is because, unlike synchronous designs, in QDI asynchronous designs, there is an intermediate RTZ or RTO phase present between two data phases. Hence, the cycle time is an important design parameter to be considered in QDI asynchronous circuits. Since a static timing analyzer normally estimates the critical path delay i.e. the forward latency, the reverse latency was estimated based on gate-level timing simulations of the asynchronous adders. Our experimentation considered a typical case PVT specification (1.05V, 25ºC) for the standard cell library [47].

About 1000 random input vectors were identically supplied to all the asynchronous adders at time intervals of 20ns through a test bench to verify their functionalities and also to capture their respective switching activities. The value change dump (.vcd) files generated through the functional simulations were used to estimate the average power dissipation. Appropriate wire loads commensurate with the different adder designs were automatically included while performing the experimentation. A virtual clock was used just to constrain the input and output ports of the asynchronous adders and it did not contribute to the area, delay or power dissipation of the adders. Table 1 presents the design metrics of the accurate and approximate 32-bit asynchronous adders, which correspond to the RTZ and RTO handshake protocols.

From Table 1, it can be noticed that the areas of the corresponding accurate and approximate asynchronous adders obeying the RTZ or the RTO protocol are the same despite having different logic compositions. In [47], the 2-input OR gate and the 2-input AND gate with a similar drive strength occupy the same area of $2.03\mu m^2$. Also, the AO22 and OA22 gates with similar drive strengths require the same area of $2.54\mu m^2$. Further, the AO222 and OA222 complex gates with similar drive strengths occupy the same area of $3.3\mu m^2$. As a result, the area occupancies of the basic building blocks, the input registers, and the completion detector for the corresponding accurate or approximate adders are the same whether they correspond to the RTZ or the RTO protocol. Having similar cell areas for the dual logic gates such as OR and AND, AO22 and OA22, and AO222 and OA222 etc. is rather uncommon in commercial standard cell libraries unlike [47]. The standard digital cell library [47] does not have foundry support and is meant for use in teaching and research. Hence, it may be hypothesized that if a commercial digital cell library is used instead for the implementation of the QDI asynchronous adders given in Table 1, then the RTO protocol would facilitate improved optimizations in the design metrics than the RTZ protocol. Therefore, the improvements in the timing and power parameters achieved by the RTO protocol than the RTZ protocol here would only serve as a baseline.

TABLE 1 DESIGN METRICS OF ACCURATE AND APPROXIMATE EARLY OUTPUT 32-BIT QDI ASYNCHRONOUS ADDERS, BASED ON A 32/28NM CMOS PROCESS

| Approximation Size | Forward Latency (ns) | Cycle Time (ns) | Area ($\mu m^2$) | Power ($\mu W$) |
|---|---|---|---|---|
| *RTZ handshake protocol* | | | | |
| Not Applicable (Accurate) | 3.02 | 3.62 | 1628.55 | 2126 |
| 4-bits (Approximate) | 2.77 | 3.37 | 1557.39 | 2124 |
| 8-bits (Approximate) | 2.44 | 3.04 | 1463.87 | 2120 |
| 12-bits (Approximate) | 2.11 | 2.71 | 1370.35 | 2117 |
| 16-bits (Approximate) | 1.77 | 2.37 | 1276.82 | 2114 |
| 20-bits (Approximate) | 1.44 | 2.04 | 1183.30 | 2110 |
| *RTO handshake protocol* | | | | |
| Not Applicable (Accurate) | 2.86 | 3.47 | 1628.55 | 2122 |
| 4-bits (Approximate) | 2.62 | 3.23 | 1557.39 | 2120 |
| 8-bits (Approximate) | 2.31 | 2.92 | 1463.87 | 2117 |
| 12-bits (Approximate) | 2.01 | 2.62 | 1370.34 | 2114 |
| 16-bits (Approximate) | 1.70 | 2.31 | 1276.82 | 2111 |
| 20-bits (Approximate) | 1.39 | 2.00 | 1183.29 | 2108 |

Although the areas of many simple and complex logic gates and their respective duals are the same in [47], when considering similar drive strengths, their delay and power dissipation values are however different. This explains why the latency, cycle time, and average power dissipation are different for the RTZ and RTO protocols, as seen in Table 1. It can also be seen in Table 1 that as the approximation size increases, the design metrics continue to decrease for the approximate asynchronous adders relative to the accurate asynchronous adders with respect to both the RTZ and RTO protocols. From Table 1, it is found that for the RTZ protocol, compared to the accurate 32-bit asynchronous adder, the approximate 32-bit asynchronous adders, on average, facilitate a 25.1% reduction in the cycle time with no trade-off in the area or average power dissipation. Likewise, for the RTO protocol, compared to the accurate 32-bit asynchronous adder, the approximate 32-bit asynchronous adders, on average, facilitate a 24.5% reduction in the cycle time at no area or power expense. In comparison with the RTZ protocol, the RTO protocol, on average, enables a 3.3% further reduction in the cycle time and leads to less power dissipation while requiring almost the same area for physical implementation.

V. Conclusions and Scope for Further Work

Accurate and approximate QDI asynchronous adders with dual-rail data encoding were realized based on RTZ and RTO handshake protocols. The adders correspond to the early output timing regime. A 32/28nm CMOS process was used as the implementation platform, and the adders were realized in semi-custom ASIC design style. The experimental results show that the approximate asynchronous adders facilitate significant reductions in all the design metrics compared to the accurate asynchronous adders. It is also noted that the RTO protocol is preferable to the RTZ protocol for the efficient realization of *approximate* asynchronous arithmetic circuits. This vindicates the observation made in [12], which inferred that the RTO protocol is preferable to the RTZ protocol for the efficient implementation of *accurate* asynchronous arithmetic circuits. Thus, this invited work, and [12], which is also an invited work together demonstrate the supremacy of the RTO protocol in effectively synthesizing accurate or approximate QDI asynchronous (arithmetic) circuits. As a possible future work, the QDI asynchronous implementation of system-level digital signal processing units could be considered to evaluate the performance of the RTZ and RTO handshake protocols.